# Nonlocal, Flat Band Meta-optics for Monolithic, High Efficiency, Compact Photodetectors

Minho Choi[1], Christopher Munley[2], Johannes E. Fröch[1,2], Rui Chen[1] and Arka Majumdar[1,2*]

[1]Department of Electrical and Computer Engineering, University of Washington, Seattle, 98195, Washington, United States.
[2]Department of Physics, University of Washington, Seattle, 98195, Washington, United States.

*Corresponding author. E-mail: arka@uw.edu;
Contributing authors: kernel@uw.edu; cmunley@uw.edu; jfroech@uw.edu; charey@uw.edu;

**Abstract**

Miniaturized photodetectors are becoming increasingly sought-after components for a range of next generation technologies, such as autonomous vehicles, integrated wearable devices, or gadgets embedded in the Internet of Things. A major challenge, however, lies in shrinking the device footprint, while maintaining high efficiency. This conundrum can be solved by realizing non-trivial relation between the energy and momentum of photons, such as dispersion-free angle-independent devices, known as flat bands. Here, we leverage flat band meta-optics to simultaneously achieve critical absorption over a wide range of incidence angles. For a monolithic silicon meta-optical photodiode, we achieved ~10-fold enhancement in the photon-to-electron conversion efficiency. Such enhancement over a large angular range of ~36° allows incoming light to be collected via a large aperture lens and focused on a compact photodiode, potentially enabling high-speed and low-light operation. Our research unveils new possibilities for creating compact and efficient optoelectronic devices with far-reaching impact on various applications, including augmented reality and light detection and ranging.

## Introduction

Absorption of light and the conversion of photons into other forms of energy, such as excitons and phonons, have been widely utilized for optical sensing[1,2], imaging[3], and energy harvesting[4]. Recently, rapid developments in wearable devices and emergence of the Internet of Things has further increased the demand for smaller photodetectors (PDs) with high absorption efficiency[5,6]. Compact integrated PDs could potentially provide several advantages for these applications, including high-resolution imaging thanks to the large pixel-count[7,8], reduced energy consumption[9], and high-speed operation[10]. However, two major limitations impede their functionalities: a small device footprint limits the total light collection, and a small device thickness limits the total amount of absorbed light (Fig. 1a). The first issue can be alleviated using a lens with large aperture to collect light from a wider area[11], while an optical cavity can circumvent the latter problem[12]. Resonant cavity-enhanced absorption occurs as the optical cavity traps light for a longer time period and by that increases the absorption efficiency. Various dielectric and plasmonic resonators have been used to enhance the photon-to-electron conversion efficiency of PDs in silicon[9,13,14] and other materials[15-18]. However, these results all focus on normal incident light, and thus is incompatible with lens-based concentrators (Fig. 1b), as focused light contains a wide range of incidence angles. Therefore, an angle-independent resonant cavity can enhance both the light-absorption and be compatible with a light concentrator to enhance the photo-responsivity (Fig. 1c).

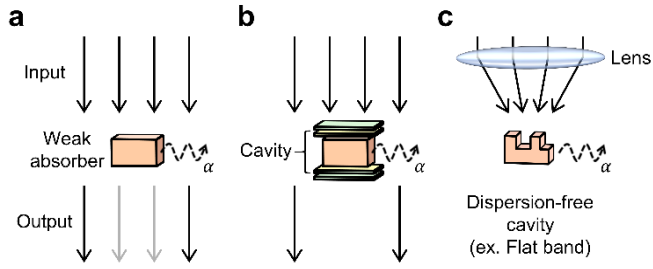

**Fig. 1 | Schematics of light absorption in a weak absorber. a**, Absorption from a weak absorber itself. **b**, Conventional cavity-enhanced absorption matching with a certain angle of incoming light. **c**, Flat band-enhanced absorption matching with a wide range of incident angle of incoming light, making it possible to use a concentrator to capture large amount of light.

The flat band concept was first conceived in condensed matter physics[19-21] and has only recently been applied to photonics in various forms[22-24]. However, to this point, photonic flat bands remained merely an interesting physical phenomenon without clear applications. Their unique property, an angle independent resonance frequency, allows cavity enhanced absorption of light consisting of many incident angles. We adapt a guided-mode resonance design in a silicon meta-optic and by fine tuning the geometrical parameter, implement a photonic flat band[25-27]. By exploiting the inherent absorption of silicon in the near-infrared, we reach critically coupled absorption with an absorption ratio ($\alpha$) of about 70% by adjusting the quality (Q) factor while maintaining the flat band. This ensures that light over a wide angular range (±18°) undergoes cavity enhanced absorption.

To harness the advantage of the enhanced absorption, we designed and fabricated a monolithic lateral p-i-n PD within the flat band meta-optics. This enables the collection of a large amount of light by using a lens, while keeping the active volume small. Thus, we achieved both light concentration using a larger aperture and cavity enhancement over a large angular range. For incident light focused with a numerical aperture (NA) of 0.13, the flat band-based PD shows a 10.3-fold enhanced responsivity compared to the unpatterned Si-PD at the resonance wavelength ~785 nm. In addition, the responsivity of this single-junction meta-optical PD depends on the wavelength and polarization of light, which can be further exploited to create a spectrally selective and polarization sensitive miniaturized PD. Such multi-functionality can reduce the fabrication costs and optical crosstalk between pixels[9], and have far-reaching impact on wearable sensors, including miniature spectrometers and Li-Fi transducers[28-31].

## Results

### Realization of photonic flat band

Cavity-enhanced absorption in conventional resonators retains a strong chromatic dependence, due to the optical path length dependence on the incidence angle. Specifically, the resonance conditions arise from constructive/destructive interference and thus a variation in the effective optical path length drastically alters it. On the contrary, photonic flat bands have no such chromatic dispersion over a certain angular range. By fine-tuning the coupling between different modes, various photonic bands including flat band, Dirac-cone, and multi-valley structures have been realized in nonlocal meta-optics[25]. By breaking the vertical symmetry through sub-wavelength patterning, such a meta-optic can also couple efficiently to the free-space[25-27]. Guided by these design insights, we created a flat band one-dimensional (1D) meta-optic in a partially-etched silicon slab on sapphire (Fig. 2a). The unit cell is composed of two symmetry broken meta-atoms. While the thickness of the silicon film is fixed at 230 nm, other parameters, including the fill factor, partial etch ratio, asymmetry ($\Delta$) between the adjacent meta-atoms, and period constitute the design space (Fig. 2b). Both the fill factor and partial etching ratio introduce a vertical symmetry breaking, while $\Delta$ and period control the Q factor and the resonance frequency, respectively.

We designed the flat band meta-optics using rigorous coupled-wave analysis (RCWA) and fabricated the structure via single-step electron beam lithography followed by reactive ion etching (Supplementary Fig. S1 and Methods). Scanning electron microscopy image of the fabricated structure is shown in Fig. 2c. Using a high-NA objective lens and high-resolution spectrometer, we set up energy-momentum spectroscopy and measured the photonic band structures (Supplementary Fig. S2 and Methods). By adjusting the fill factor of the structure, we realized various photonic band structures including Dirac-cone and photonic flat band as shown in Fig. 2d and Supplementary Fig. S3. Noticeably, the flat band emerges with an isofrequency resonance in the near-infrared region (~780 nm) and is maintained over a wide angular range of ±18°. In comparison other resonators show a strong dependence of the resonance frequency on the angle of incidence.

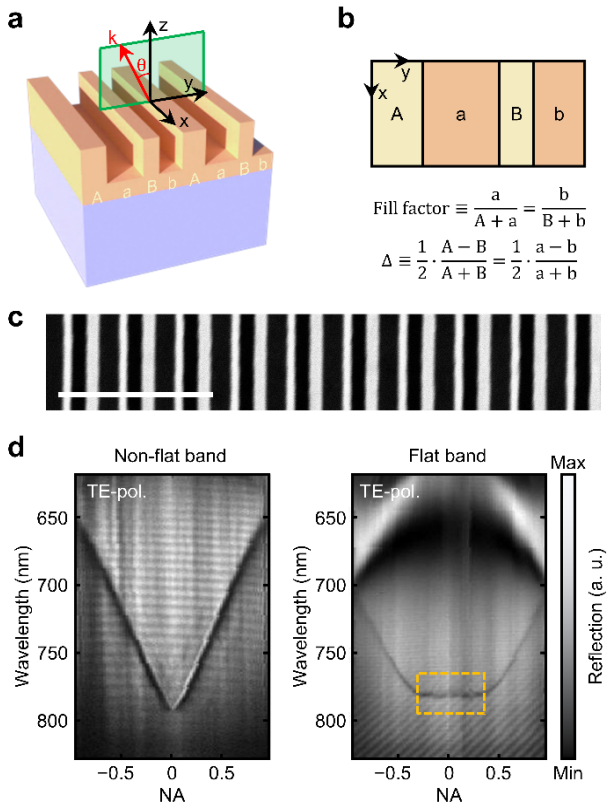

**Fig. 2 | Implementation of photonic flat band as an angle-independent resonator. a**, Schematic of a flat band meta-optic which has a wide range of angle, θ, independent response. **b**, Top view of the meta-optics consisted of four different parts due to the partial etching, and definitions of fill factor and asymmetry factor, Δ. **c**, Scanning electron microscope image of the flat band meta-optic. The scale bar: 1.0 μm. **d**, Reflective energy-momentum spectra of the meta-optics: Dirac-cone and flat band, respectively. A yellow dashed box indicates the flat band.

**Photonic flat band and critically-coupled absorption**

To understand how the photonic flat band can be leveraged to maximize the absorption efficiency, we consider a configuration of two partially reflecting mirrors and a weak thin-film absorber in between (Fig. 3a). In this analogy, the flat band meta-optic is equivalent to a curved cavity, which facilitates cavity enhanced absorption simultaneously for a wide angular range of incident light (Fig. 3b). The boundary conditions, however, to maximize the absorption inside the cavity are not trivial, and simply maximizing the Q-factor or increasing the absorption coefficient of the device would not directly maximize the absorption. Therefore, to predict whether the flat band meta-optic exhibits critically-coupled absorption, we modeled the system behavior using the finite-difference time-domain (FDTD) method (see details in

Methods). We studied both the reflective and transmissive response of the meta-optic using a model complex refractive index, $3.7 + i\kappa$, for the crystalline silicon. For a zero-absorption coefficient, i.e., $\kappa = 0$, we clearly observe a Fano-like resonance from both reflection and transmission spectra, and no absorption occurs at any wavelength (Fig. 3c). On the other hand, if there is a finite value of $\kappa$, the Fano-like resonance becomes weaker, and clear cavity-assisted absorption occurs (Fig. 3d). As we gradually increase $\kappa$, from 0 to 0.1, the meta-optic meets the critically coupled absorption point at $\kappa \sim 0.005$ with $\alpha$ of about 70% which is much higher than a GMR structure with higher $\kappa$ of 0.1 where $\alpha$ is about 25% (Fig. 3E and Supplementary Fig. S5a).

In practice, $\kappa$ is an intrinsic property and has a fixed value, thus other factors have to be adjusted to reach the critical absorption condition. Fortunately, we can fine tune the cavity Q-factor by adjusting the asymmetry, $\Delta$, of the structure – as the value of $\Delta$ reduces, the resonance becomes narrower, similar to a bound-state-in-continuum resonance[26]. From FDTD simulations with the measured refractive index, we can obtain a Q factor of about 174 at $\Delta \sim 0$, while the Q factor reduces gradually as $\Delta$ increases and reaches ~50 for $\Delta = 0.26$ (Supplementary Fig. S5b). By gradually changing the $\Delta$ of the structure from 0 to 0.35 in simulation, we can find the critically-coupled absorption point near $\Delta = 0.20$ where $\alpha$ is about 70%, consistent to the simulated result of varying $\kappa$ (Fig. 3f). The measured results are consistent with the simulation, as we can see the change in the cavity resonance and linewidth as $\Delta$ increases from 0.05 to 0.20 (Supplementary Fig. S5d). We measured both reflection (R) and transmission (T) spectra of the meta-optics with varying $\Delta$, and the sum of normalized reflection and transmission is shown in Fig. 3g. We measure $\alpha \sim 68\%$ at the critical coupling condition, which is similar to the simulated value of ~70%. Theoretical explanations are shown in Supplementary Information II.

We note that ~60 μm-thick silicon and antireflection coating is needed to achieve 70% absorption without cavity enhancement at ~750 nm wavelength, where silicon has the absorption coefficient of around 0.014. Such a thick medium hinders a fabrication availability and also reduce the operation speed of the device due to a long-distance carriers have to travel. But with the cavity enhancement, only a 230-thick silicon can achieve the same amount of absorption. On the other hand, we can tune the resonant wavelength of the flat band meta-optic by adjusting the structural parameters. In detail, silicon generally has less absorption coefficient at longer wavelengths; then the Q factor of the structure needs to be enhanced which corresponds to a decrease in $\Delta$ (Supplementary Information II).

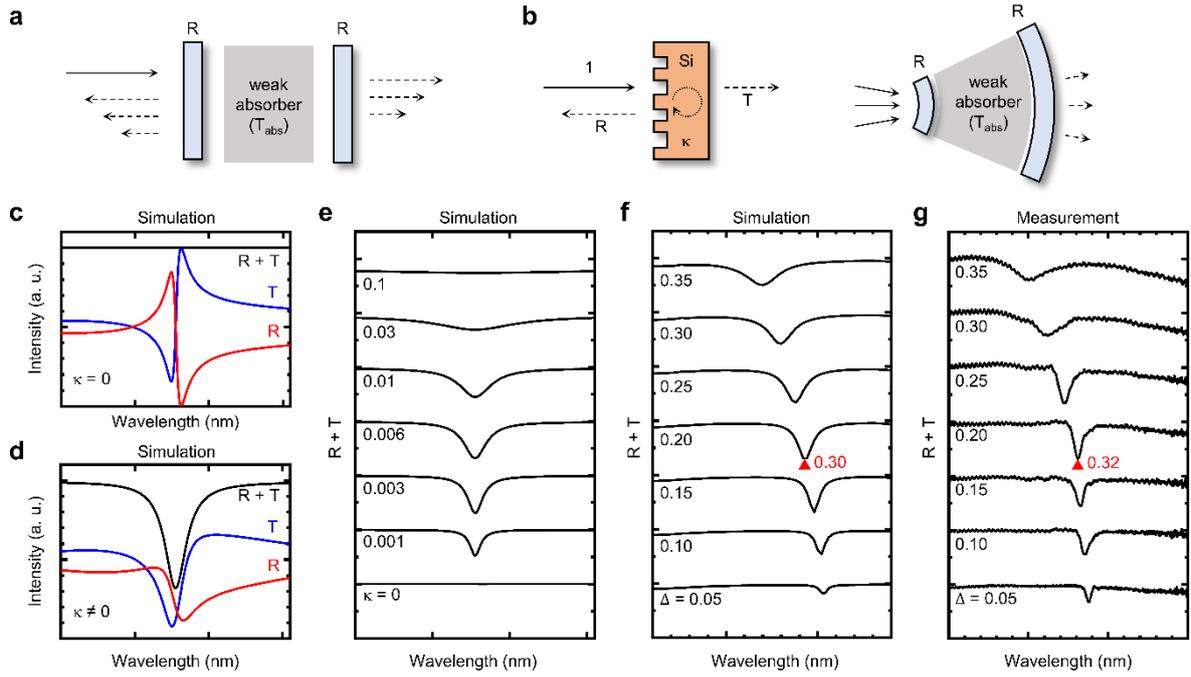

**Fig. 3 | Flat band meta-optic and critically-coupled absorption. a**, Conventional optical resonator with a weak absorber inside. **b**, Interpretation of the flat band meta-optics as a cavity and a weak absorber. **c,d**, Simulated reflection, R, and transmission, T, spectra of the flat band meta-optics without and with absorption, respectively. **e,f**, Simulated absorption spectra with respect to κ and Δ, respectively. **g**, Experimentally measured absorption spectra with respect to Δ. Here, we used following values for the meta-optics parameters (fill factor, partial etching ratio, and period are 0.78, 0.84, and 384 nm, respectively), while varying the Δ.

**High-performance PD with flat band meta-optic**

We demonstrate a compact, high-responsivity PD using the optimized silicon flat band meta-optic, exploiting the critically-coupled absorption over a wide angular range of incident light. This enables us to significantly increase the total light collection as compared to other cavity-enhanced PDs[9,13-18]. As shown in Fig. 4a, a lens on top of a compact, single-junction PD can be used to increase the amount of light collected and absorbed into a single PD.

Using an additional lithography step, we selectively doped the silicon meta-optic and created a lateral p-i-n PD with 10-μm-wide intrinsic region. Fig. 4b shows images of multiple PDs connected to metal pads and wires (Supplementary Fig. S6 and Methods). To rule out variations and to clarify the effect of flat band meta-optics compared to others, i.e. unpatterned silicon films and "non-flat band" meta-optic, we fabricated multiple devices in a row and put a pair of large metal pads to connect the p- and n-doped regions of all devices. In a single chip, 14 devices were connected with large metal pads at the same time. Fig. 4c shows an I-V characteristic of the entire chip from −5 V to 5 V, with less than 50 nA of leakage current at reverse bias voltage and a clear rectification behavior of the diode. Considering the number of devices on the chip, we can estimate that each single device has a leakage current of less than 3 nA at a reverse bias voltage of −5 V or less.

We focused light with an NA of 0.13 to an intrinsic region of the device and measured the photocurrent, which is linearly proportional to the optical power of the incident light, at a reverse bias voltage of −2 V (Fig. 4c). Here, the slope of the linearly fitted line gives us the responsivity (η) of the PD, which represents the conversion efficiency from photon energy into electrical current. We note that this monolithic silicon PD could generate photocurrent from a few tens of nanowatt light. An experimental setup for the photocurrent measurement is shown in Supplementary Fig. S7.

We then measured a series of power-dependent photocurrents to extract η while adjusting the polarization and laser wavelength from 765 to 799 nm in 2 nm steps. The photocurrent measurement was repeated five times to obtain an average value. As shown in Fig. 4d, the average values of η for a 230-nm-thick unpatterned silicon film are 3.2 ± 0.2 (mA/W) and 3.4 ± 0.1 (mA/W) for x- and y-polarized light along the wavelength from 765 to 799 nm, respectively. There was no clear trend in wavelength or polarization direction. On the contrary, we can observe the wavelength and polarization dependence in η for a meta-optic with "non-flat band" photonic structure (Fig. 4e). The average values of η are 5.7 ± 0.4 (mA/W) and 0.5 ± 0.1 (mA/W) for x- and y-polarized light along the 765 to 799 nm wavelength, respectively. It shows an enhanced η for x-polarized light at a wide range of wavelengths from 773 to 795 nm and had a maximum value of η = 10 ± 2 (mA/W) at 789 nm wavelength. However, this enhancement occurred only for x-polarized light, while even a reduction occurred for y-polarized light compared to the unpatterned silicon film because the meta-optic has less absorptive material, after the patterned etching.

As shown in Fig. 4f, for the flat band meta-optic, we observed a much stronger and sharper enhancement of η as the resonance wavelength remains the same over a wide range of incident angles. The average values of η for the HCG structure with flat band are 15 ± 0.7 (mA/W) and 3.1 ± 0.1 (mA/W) for x- and y-polarized light along the 765 to 799 nm wavelength, respectively. It shows a strongly enhanced η for x-pol at wavelengths near 785 nm, with a maximum value of η = 35 ± 1 (mA/W) at 785 nm wavelength. All the measured data for flat band-based PD is shown in Supplementary Fig. S8. The resonance wavelength of ~785 nm is well-matched with the measured photonic band structure of the flat band (Supplementary Fig. S9). We note that the higher η values compared to the non-flat band meta-optic, occur due to the lower fill factor of the flat band structure and therefore more absorptive material.

Overall, we achieve the enhancement in η for the flat band-based PD, given as $\eta_{flat}/\eta_{film}$ = 10.3 ± 1.6 and $\eta_{flat}/\eta_{non-flat}$ = 3.5 ± 0.7 in comparison to the unpatterned silicon film and the non-flat band meta-optic, respectively. The flat band-based PD has a highly-dependent η with respect to both wavelength and polarization, where the full-width half-maximum of the wavelength-enhancement region is ~10 nm and the extinction ratio of orthogonal, linear polarization is 0.86 ± 0.09. From FDTD simulations, the Q factor of the meta-optic is around 76 at Δ = 0.20 (Supplementary Fig. S5b) which matches well with the enhanced response bandwidth of the PD. In this regards, the flat band-based PD can be considered an integrated PD with a wavelength selective filter and linear polarization sensitivity, essentially combining three functionalities into a single meta-optic, free of complex fabrication and optical crosstalk between optics[9].

Moreover, we simulated the NA-dependent absorption from the 230-nm-thick unpatterned silicon film as a reference to extract the absorption enhancement factor (Z) of the flat band meta-optic. Here, Z is a term that only describes the light absorption but can be compared to the PD efficiency enhancement factor, $\eta_{flat}/\eta_{film}$. This is because the amount of light absorption is proportional to the generated free carrier density which corresponds to the photocurrent amplitude under a reverse bias voltage. Since all the PDs, including the unpatterned silicon film, non-flat band meta-optic, and flat band meta-optic, share the same electrical components (i.e. contact pads and voltage source) we can assume that the converting efficiency from free carriers to photocurrents is the same for all PDs. At an NA of 0.13, the simulated Z and experimentally measured $\eta_{flat}/\eta_{film}$ are in good agreement, Z = 10.4 and $\eta_{flat}/\eta_{film}$ = 10.3 ± 1.6 (Supplementary Fig. S10c).

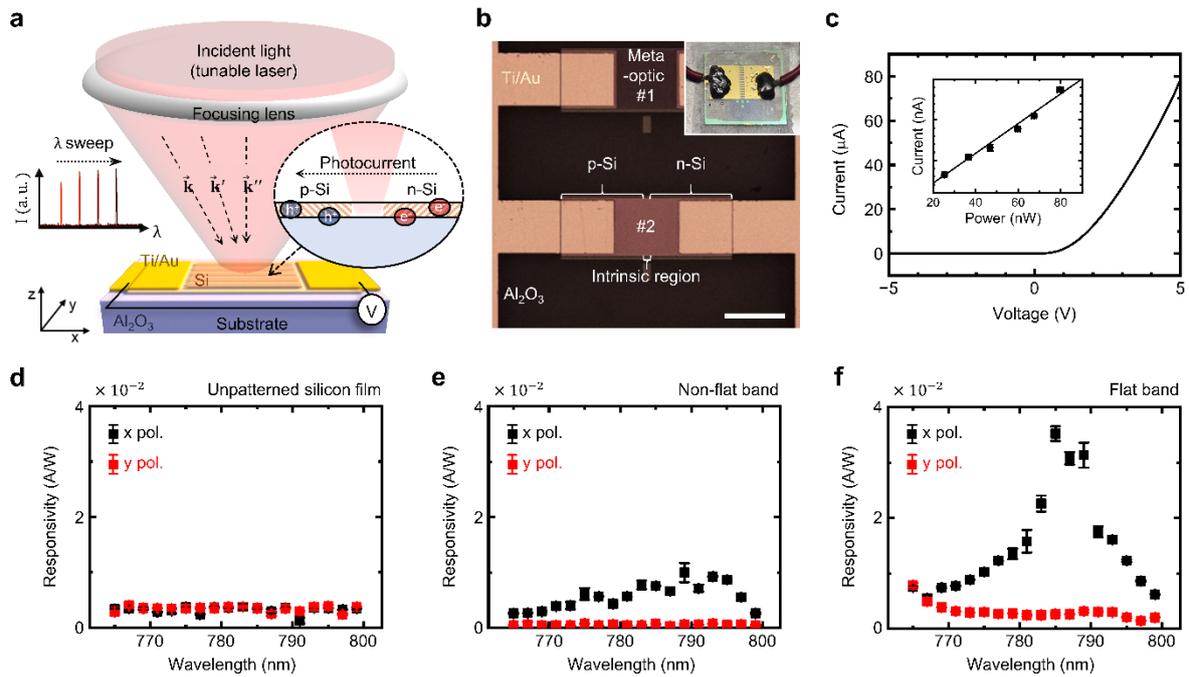

**Fig. 4 | Lateral p-i-n PD with flat band meta-optic**. **a**, Schematics of the lateral p-i-n PD using flat band meta-optic. Large amount of light is collected and focused onto a compact flat band-based PD by a lens. A tunable laser allows wavelength-selective response measurement of devices. **b**, Optical microscope image of the device. The inset image shows the actual device after soldering wires to the metal pads. The length of the scale bar is 200 μm. **c**, I-V curve of the PD without incident light. The inset graph shows the current as a function of incident light power. The slope of the linear fit indicates the responsivity, η, of the device. **d**-**f**, Wavelength- and polarization (pol.)-dependent η of the PDs made of (d) 230-nm-thick unpatterned silicon film, (e) Meta-optic with "non-flat band" photonic structure, and (f) Meta-optic with photonic flat band structure. The photonic band structure of "non-flat band" is represented in Fig. 2d.

## Discussion

The measured responsivity, η, of the flat band meta-optical PD is about 0.04 A/W at its resonance wavelength of 785 nm, corresponding to a quantum efficiency (QE) of about 6%. Although this value is smaller than current state-of-the-art commercial silicon PDs[32], which have QE of ~95%, and widely-used commercial CMOS camera, i.e. Thorlabs, CS165MU, which has QE of ~35% in the near-infrared region; our flat band-based PD has an order of magnitude smaller thickness. This could potentially provide advantages in terms of supply voltage, dark current, and response speed. We envision that this value could be further boosted by further optimization of the p-i-n junction parameters (e.g. active layer width, doping concentration, and annealing temperature)[33,34], or by adding a photoconductive gain with CMOS compatible process[35]. Nevertheless, our approach demonstrates how a photonic flat band can be exploited to enhance the light absorption for focused, incident light, as we achieve a 10.3-fold enhancement from the flat band-based PD for focused light of NA 0.13 compared to an unpatterned silicon film-based PD.

In addition, our flat band meta-optical PD provides a wavelength- and polarization-selective response, both of which are adjustable through the specific design parameters. The linear polarization selectivity of the device can be modulated by the device orientation, and the wavelength selectivity can be adjusted by the period of the structure or changing the effective refractive index of the structure (e.g., by integrating phase-change materials with the meta-optics)[36,37]. By tuning the resonance frequency of the flat band meta-optic spectral information of the incident light can be collected by a single PD, which is equivalent to a compact spectrometer. In case of conventional photonic band structures rather than a photonic flat band, the resonance frequency varies greatly depending on the incident angle of light; therefore, it is difficult to confirm the wavelength of measured light unless the incoming light has a certain incident angle, e.g. normal incidence. In comparison, the photonic flat band maintains its resonance frequency irrespective of the incident angle.

## Conclusion

We realized cavity-enhanced absorption using a single flat band meta-optics with absorption power of 68% for wide angular range of NA 0.35, which is about 25 times more than the previous angular resolution of angle-independent cavities using bulk optics[38]. We utilized this silicon flat band meta-optic as an efficient and compact PD. This flat band-based PD can provide improved absorption over a wide range of angles of incidence, allowing light collection with a large aperture in small footprint PD. Due to the 1D characteristics of the flat band, it has a particular k-vector and linear polarization dependence, which results in NA-dependent absorption enhancement. Hence, we can consider this flat band-based PD as a miniaturized PD which has wavelength and polarization filters insertion. This work, which is not limited to silicon materials, showcases the first practical application of photonic flat band as a degenerate resonant cavity-enhanced absorber, with potential impacts in fields requiring wavelength-selective compact detectors or angle-insensitive detectors.

## Methods

### Numerical simulation

A commercial software Ansys Lumerical RCWA was used to design a photonic flat band meta-optic. The meta-optics has a comb structure implemented in a silicon on sapphire wafer. For the RCWA simulation, we used refractive indexes of 3.7 and 1.761 for silicon and sapphire, respectively. To simulate the absorption spectra of the meta-optic, either flat band or bullseye structure, another commercial software Ansys Lumerical FDTD was used. We used plane-wave and Gaussian source on the meta-optic and extracted the absorption for normally-incident or focused light irradiation on the meta-optic. For the FDTD simulation, we used the refractive index data measured by an ellipsometer (Woollam M-2000) for the silicon film.

### Device fabrication

A commercial 230 nm thick silicon film on sapphire wafer (UniversityWafer) was used for the flat band meta-optic. First, the wafer was cleaned with acetone and isopropyl alcohol and then $O_2$ plasma ashing for 0.2 min at 100 W. Then, a positive-tone E-beam resist (ZEP-520A) was coated and bake for 3 min at 180°C followed by coating a charge-dissipation layer (DisCharge $H_2O$). After the E-beam lithography (JOEL JBX6300FS), we developed the resist with amyl acetate for 2 min. Then, we clarify the pattern with another O2 plasma ashing process for 0.1 min at 100 W, then etch the silicon layer with plasma etching (Oxford PlasmaLab 100) with fluorine etch gases of $SF_6/C_4F_8$. At last, we dissolved the resist with n-methyl propanol. Then, two cycles of photolithography and ion implantation processes for selective p- and n-type doping on each half region of the flat band meta-optic, except the central 10-μm-wide intrinsic area. We coated hexamethyldisilane as an adhesive layer and then a positive-tone photoresist (AZ1512), followed by baking for 1 min at 100°C. We patterned the aperture for p- (n-) doping region of the silicon film by direct laser write lithography (Heidelberg DWL-66+), and developed the resist with AD-10 for 1 min after the post-bake process for 1 min at 100°C. Then, we implanted boron (phosphorus) ions with dose of $2\times10^{15}$ ions per $cm^2$ and energy of 14 (40) keV. Then we annealed to activate both p- and n-type ion implantations at 950°C for 10 min (Expertech CRT200 Anneal Furnace) for dopant activation. Then we isolated each device by fully etching the silicon film excluding the device because a silicon film can become a current leakage channel even it is not intentionally doped We did direct laser write lithography followed by plasma etching to eliminate all the silicon except the p-i-n devices and markers for alignment. Then, we deposited a pair of Ti/Au metal pads of 10/150 nm-thickness on each of the p- and n-doped regions (CHA SEC-600), and connected each pad to an external modulator with soldering wires for electrical modulation.

### Experimental set-up

We built an energy-momentum spectroscopy setup to measure photonic band structure of the meta-optic. We used a high-NA objective lens (Olympus MPLAPON100X) to collect large amount of angular information of the photonic band structures. From the normal micro-photoluminescence measurement setup, we added a Fourier-lens in between the objective lens and the last lens which collecting the light into the spectrometer (Teledyne IsoPlane), so that the k-space information rather than real-space was collected on the detector. Then we spatially filtered with slit on the spectrometer, and spectrally filtered with grating. We used the objective lens with NA 0.95 to collect the reflected or transmitted light while using the same objective lens or another objective lens (Olympus LMPLFLN50X) to focus the broad-band light (Xenon lamp) onto the meta-optics surface for measuring the energy-momentum spectrum in reflection or transmission configuration. We used 100,000:1 polarizer (Thorlabs Glan-Taylor) to select the TE-polarization for energy-momentum spectroscopy. For the PD characterization, we use a tunable laser (Toptica DL pro) to adjust the input laser wavelength and focus the light onto the meta-optic-based PD. We used a 50:50 beam splitter to divide the incident laser, one to the highly-sensitive power meter

(Thorlabs, S130C) and the other one to the device, and measured the power-dependent photocurrent under −2V bias voltage using a source measure unit (Keithley 2450 SourceMeter). We adjusted the polarization using the 100,000:1 polarizer, and swept the laser wavelength to achieve the polarization- and wavelength-dependent response of the flat band-based PD.

## Data availability

All data needed to evaluate the conclusions in the paper are present in the paper and/or the Supplementary Information. The data that support other findings of this study are available from the corresponding author on request.

## Acknowledgements

This material is based upon work supported by the National Science Foundation Grant No. DMR-2019444. Part of this work was conducted at the Washington Nanofabrication Facility / Molecular Analysis Facility, a National Nanotechnology Coordinated Infrastructure (NNCI) site at the University of Washington with partial support from the National Science Foundation via awards NNCI-1542101 and NNCI-2025489.

## Author contributions

M.C. and A.M. conceptualized the project. M.C. and C.M. designed, M.C., J.E.F. and R.C. fabricated, and M.C. measured the devices. M.C. and C.M. analyzed the data. M.H. visualized the data. A.M. supervised the project. M.C. wrote the original draft of the manuscript, and all the authors worked on review and editing.

## Competing interest

The authors declare that they have no competing interests.

# Supplementary Information: Nonlocal, Flat Band Meta-optics for Monolithic, High Efficiency, Compact Photodetectors

Minho Choi, Christopher Munley, Johannes E. Fröch, Rui Chen, and Arka Majumdar

**CONTENTS**

# I. Realization of photonic flat band with meta-optics

## 1.1. Rigorous coupled-wave analysis (RCWA)

We realized photonic flat bands with one-dimensional (1D) meta-optics of crystalline silicon on a sapphire wafer. Using a broadband transparent sapphire substrate, absorption of the meta-optics can be estimated by measuring both the reflection and transmission. We are making "comb" structures to create photonic flat bands based on vertical symmetry breaking. This symmetry breaking causes coupling different modes and results in various photonic structures, e.g. parabola, flat band, and multi-valley [S1]. Various approaches have been implemented to create photonic flat bands [S2–S4], but the comb structure has its advantages of simple, one step fabrication and a large degree of freedom to modulate the photonic band structures. We have four independent parameters of fill factor, partial etching ratio, asymmetry factor ($\Delta$), and period to define the comb structure, and both the fill factor and partial etching ratio are related to the vertical asymmetry of the structure. Therefore, we could tune photonic band structures by adjusting either of those two factors, and indeed found photonic flat bands by simulating these designs with RCWA. We should note that the partial etching ratio is one of the most important factors, since there will be no vertical symmetry breaking without partial etching. In the reference, Nguyen et al., an etch stop layer of $SiO_2$ was used to control the partial etching depth of the structure [S1]. On the contrary, we controlled the partial etching depth with precise control of etching time and etching speed. With our plasma etching recipe the rate is about 3.28 nm/sec, and etching time is around 58 sec for partially etching ~ 190 nm depth from 230-nm-thick silicon slab. But still, it is difficult to expect absolutely consistent etching speed all the time due to the slight fluctuation of gas flows, pressure, and power. In order to compensate for the uncertainty of partial etching depth, we patterned multiple meta-optics with a series of fill factors so that one of the meta-optics would match the etching depth. According to the RCWA simulation, the photonic flat band occurs at a larger fill factor when the partial etching depth is larger, while the resonance frequency blue shifts (Supplementary Fig. S1).

## 1.2. Optical characterization of photonic band structure

We made an energy-momentum spectroscopy setup with a high-numerical aperture (NA) objective lens and a high-resolution spectrometer (Supplementary Fig. S2). With a detachable mirror, we could choose either a reflection or a transmission configuration for energy-momentum spectroscopy. For the reflection configuration, an objective lens with an NA of 0.95 objective lens was used for both input and output light from the meta-optic; however, an objective lens with an NA of 0.55 (0.95) objective lens was used for input (output) for the transmission configuration due to the short working distance (~0.35 mm) of the objective lens with an NA of 0.95 making it difficult to put the meta-optic in between two high-NA objective lenses. Hence, the energy-momentum spectra obtained from the transmission configuration have less momentum information than those from the reflection configuration. We clearly observed not only the photonic flat band, and also the evolution of photonic band structures from parabola to flat band and multivalley by adjusting the fill factor of the meta-optic, analogue to the RCWA simulation (Supplementary Fig. S3a). The flat band is observed in the transmission configuration as well, while a range NA is more limited compared to the reflection configuration due to the smaller NA of the objective lens in the input (Supplementary Fig. S3b).

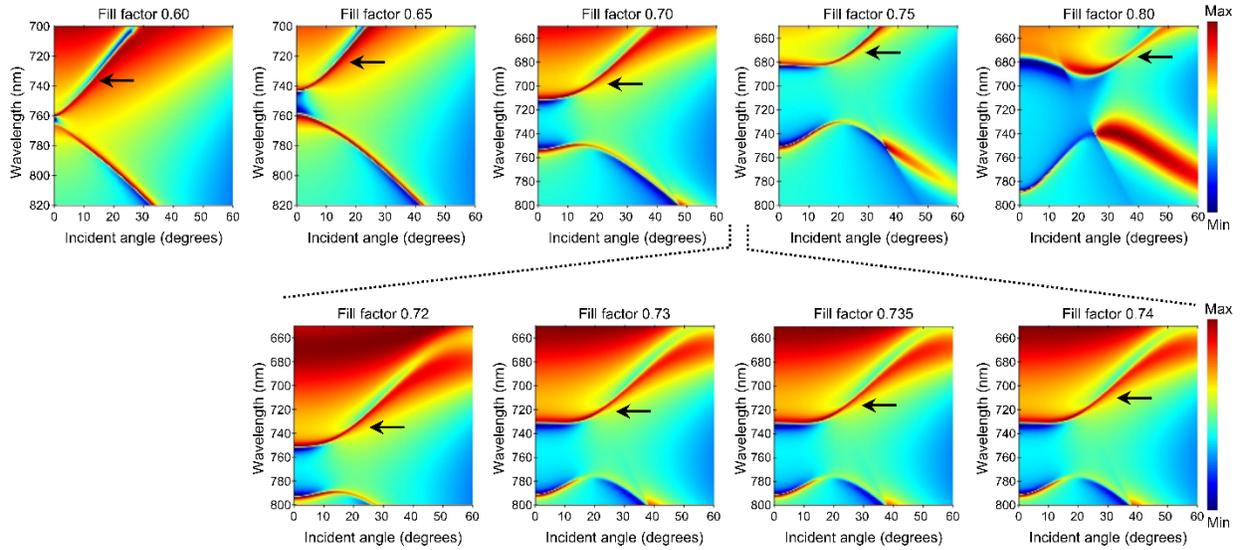

**Supplementary Figure S1. RCWA simulation of angle-dependent transmission spectra in meta-optics.** Angle-dependent transmission spectra of the meta-optics for TE-mode polarized light were calculated by RCWA simulation. The meta-optics consist of a comb structure of crystalline silicon on sapphire substrate, and we use refractive indices of 3.7 and 1.761 for silicon and sapphire, respectively, in the RCWA simulation. We varied the fill factor of the structure while fixing other defining parameters of the structure as following: partial etching ratio = 0.80, $\Delta$ = 0.15, period = 320 nm. Black arrows indicate the photonic band that we are focusing on, and it is clear that photonic band structure evolves from Dirac-cone to flat band and multivalley as the fill factor of the meta-optic increases.

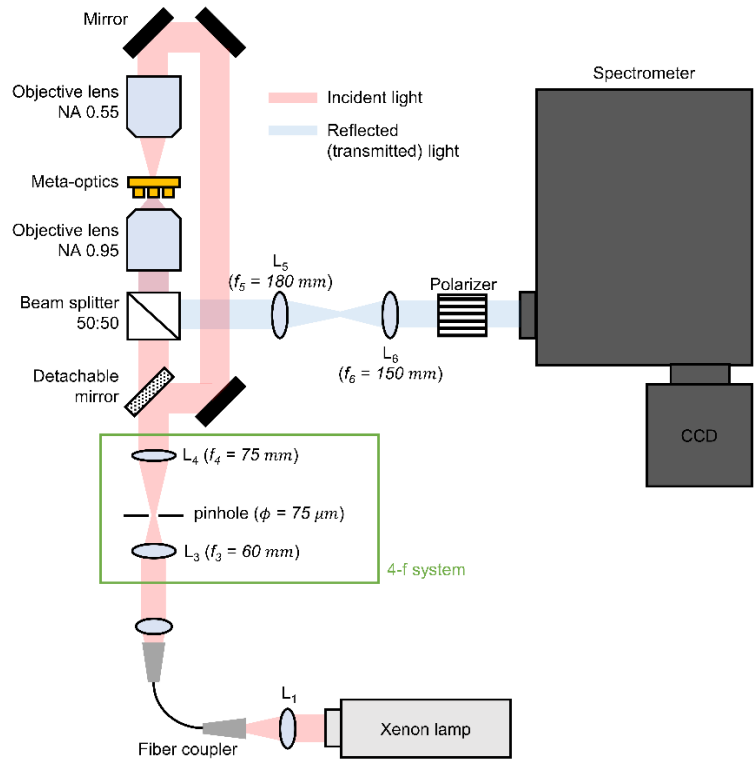

**Supplementary Figure S2. Schematic of the experimental setup for the energy-momentum spectrum measurement.** Broad-band Xenon lamp is used to characterize reflection (transmission) energy-momentum spectra of meta-optics. A 4-f system is used to finely collimate the light and enlarge the beam size. A detachable mirror was used to measure in either the reflection or transmission configuration. A high-NA objective lens (NA = 0.95) is used for both focusing and collecting light in the reflection configuration, but a low-NA objective lens (NA = 0.55) is used only for focusing light in the transmission configuration, while a high-NA objective lens is used for collection. This is because the working distance of the high-NA objective lens is too short, ~ 0.35 mm, and we were unable to insert the meta-optic when using two high-NA objective lenses. The Fourier lens (L5) allows the detector to distinguish angular information instead of spatial information in both reflection and transmission, and the grating inside the spectrometer allows the detector to distinguish spectral information.

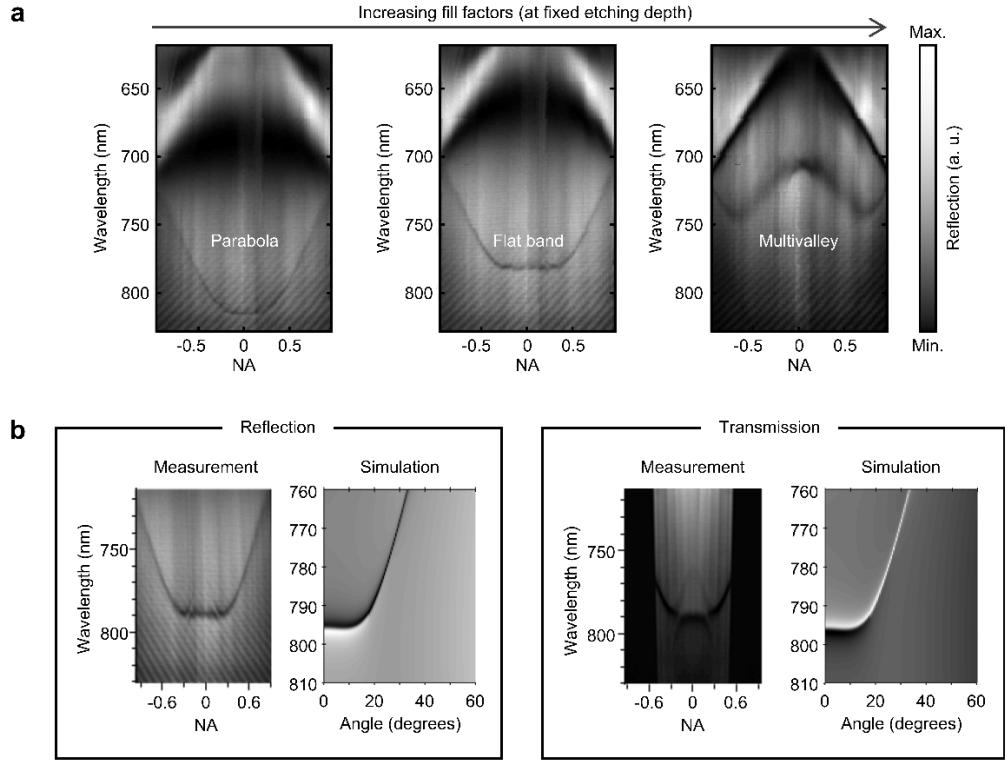

**Supplementary Figure S3. Energy-momentum spectra of the meta-optics. a**, Measured energy-momentum spectra of the meta-optics with varying fill factors. The phonic band structures are tuned from parabola to flat band, and multivalley. **b**, Measured and simulated photonic flat band in both reflection and transmission configuration, respectively. Photonic flat band has been absorbed in both reflection and transmission configuration, while the angular range of the energy-momentum setup in transmission configuration is limited compared to the reflection configuration due to the low-NA objective lens for input light (which is necessary due to too short working distance of high-NA objective lens).

## II. Critically coupled absorption and photonic flat band

### 2.1. Critical coupling: A weak absorber in between a resonant cavity

The silicon itself has an intrinsic absorption coefficient for above-bandgap light, which corresponds to a wavelength of 1100 nm. Due to the indirect bandgap nature of silicon, it does not have significant absorption compared to the other direct band gap materials. Still, in some cases, if even a small amount of absorption needs to be avoided, then larger bandgap material, e.g. SiN, SiC, or GaP, are used to minimize the optical losses within the material [S5]. However, on the contrary, we maximized the inherent absorption of silicon material with resonant cavity integration and achieved an absorption ratio ($\alpha_c$) of ~70% in the near-infrared region. Critical coupling, which maximizes the absorption, occurs at the flat band. We can explain how the critically coupled absorption occurs by modeling the silicon meta-optic as a weak absorber and two partially reflective mirrors sandwiching the weak absorber. For analytical calculation, let us define $r_1$, $r_2$, $T_{abs}$, and $\phi$ as the reflectivity of the first mirror where the light comes in, reflectivity of the second mirror, transmittance of the weak absorber, and phase delay for passing the weak absorber.

Then, the reflection or transmission amplitude of the light with respect to the incidence angle can be described as an infinite series of terms describing how many times the light reflected off of the mirrors:

$$\frac{r}{i} = -r_1 + (1-r_1)^2 r_2 T_{abs}^2 e^{i2\phi} + (1-r_1)^2 r_1 r_2^2 T_{abs}^4 e^{i4\phi} + \cdots = -r_1 + \frac{(1-r_1)^2 r_2 T_{abs}^2 e^{i2\phi}}{1-r_1 r_2 T_{abs}^2 e^{i2\phi}},$$

$$\frac{t}{i} = (1-r_1)(1-r_2)T_{abs}e^{i\phi} + (1-r_1)(1-r_2)r_1 r_2 T_{abs}^3 e^{i3\phi} + \cdots = \frac{(1-r_1)(1-r_2)T_{abs}e^{i\phi}}{1-r_1 r_2 T_{abs}^2 e^{i2\phi}}.$$

Here, we put additional π-phase for the reflection when light incidents from outside of the pair of mirrors, considering the high refractive medium (e.g. silicon) is filled between the mirrors. In the case of Supplementary Fig. S4a, where the second mirror has a perfect reflectivity ($r_2 = 1$), both reflection and transmission amplitude can be described as following:

$$\left.\frac{r}{i}\right|_{r_2=1} = -r_1 + \frac{(1-r_1)^2 T_{abs}^2 e^{i2\phi}}{1-r_1 T_{abs}^2 e^{i2\phi}},$$

$$\left.\frac{t}{i}\right|_{r_2=1} = 0.$$

It is obvious that there is no transmitted light as the second mirror has perfect reflectivity. In this case, we can find a condition where there is no reflected light as well, which means all the incident light is absorbed (Supplementary Fig. S4b). This system is called a coherent perfect absorber (CPA), and in this particular case, a self-interfering CPA [S6–S8]. The optimized reflectivity of the first mirror ($r_1^*$) condition according to the given $T_{abs}$ for self-interfering CPA is as following:

$$r_1^* = \frac{1}{4}\left(\frac{1}{T_{abs}^2} - \sqrt{\frac{1}{T_{abs}^4} + \frac{4}{T_{abs}^2} - 4} + 2\right),$$

$$e^{i2\phi} = 1.$$

Supplementary Fig. S4c shows the positive tendency between $r_1$ and $T_{abs}$ at the CPA condition. The positive relationship between two represents that we need higher quality (Q) cavity for weaker absorber. But a critically coupled absorption condition always exists for any $T_{abs}$.

In case of Supplementary Fig. S4d, where the second mirror has imperfect reflectivity but has the same reflectivity as the first mirror ($r_2 = r_1 < 1$), assuming the system is symmetric, then both reflection and transmission amplitude can be described as following:

$$\frac{r}{i} = -r_1 + \frac{(1-r_1)^2 r_1 T_{abs}^2 e^{i2\phi}}{1-r_1^2 T_{abs}^2 e^{i2\phi}},$$

$$\frac{t}{i} = \frac{(1-r_1)^2 T_{abs} e^{i\phi}}{1-r_1^2 T_{abs}^2 e^{i2\phi}},$$

also considering the additional π-phase shift for the light incident from outside of the pair of mirrors. For the constructive interference condition ($e^{i2\phi} = 1$), the reflection and transmission light intensities ($I_R$ and $I_T$) with respect to the incident intensity (I) can be described as following:

$$\frac{I_R}{I} = \left(-r_1 + \frac{(1-r_1)^2 r_1 T_{abs}^2}{1 - r_1^2 T_{abs}^2}\right)^2 = r_1^2 - \frac{2r_1^2(1-r_1)^2 T_{abs}^2}{1 - r_1^2 T_{abs}^2} + \frac{(1-r_1)^4 r_1^2 T_{abs}^4}{(1 - r_1^2 T_{abs}^2)^2},$$

$$\frac{I_T}{I} = \frac{(1-r_1)^4 T_{abs}^2}{(1 - r_1^2 T_{abs}^2)^2}.$$

In this case, we cannot make both the reflection and transmission power to be zero at the same time which means that there always some amount of light that is not absorbed by the medium. But still, we can maximize the absorption under the critical coupling condition. The amount of light absorption ($I_{abs}$) with respect to the incident light can be described as following:

$$\frac{I_{abs}}{I} = 1 - \frac{I_R}{I} - \frac{I_T}{I} = 1 - r_1^2 + \frac{2r_1^2(1-r_1)^2 T_{abs}^2}{1 - r_1^2 T_{abs}^2} - \frac{(1-r_1)^4 (r_1^2 T_{abs}^2 + 1) T_{abs}^2}{(1 - r_1^2 T_{abs}^2)^2}.$$

Supplementary Fig. S4e shows the $I_{abs}$ with respect to $r_1$ at three different $T_{abs}$ values, 0.4, 0.6, and 0.95. We can find the optimized $r_1^*$ for critically coupled absorption, which maximizes the absorption, for each $T_{abs}$ values. Here, the optimized $r_1^*$ value for critical coupling increases as $T_{abs}$ increases as same as the previous case where $r_2 = 1$, while the maximized $I_{abs}$ we can achieve decreases. This trend can be brought into the discussion of the Q factor and the absorption coefficient (κ) of the silicon meta-optics. As the κ increases, $T_{abs}$ decreases, then optimized $r_1^*$ for critical coupling decreases, which corresponds to the decreasing the optimized Q factor of the cavity for maximizing the absorption.

$$\kappa \uparrow \Rightarrow T_{abs} \downarrow \Rightarrow r_1^* \downarrow \Rightarrow Q^* \downarrow$$

2.2. Demonstration of critically-coupled absorption with silicon, flat band meta-optics

For the flat band meta-optic made of silicon on a sapphire substrate, we used finite-difference time-domain (FDTD) simulation to calculate the absorption power. We could extract the amount of absorption by subtracting both reflected and transmitted light from the incident light. To achieve the critically-coupled absorption for the flat band meta-optics, we first put a model refractive index, $3.7 + i\kappa$, for the silicon material and simulated the absorption spectra depending on κ. We could achieve the maximum absorption of ~70% when κ was 0.003 (Supplementary Fig. S5a). However, in practice, κ is an intrinsic property of the material which varies with the wavelength; therefore, we cannot adjust it (which corresponds to the $T_{abs}$). Fortunately, we can adjust the Q factor of the meta-optics quite easily as the Δ is closely related to the Q factor (which corresponds to the $r_1$), similar to the bound states in continuum (Supplementary Fig. S5b) [S9]. Hence, we could find the critically coupled absorption by adjusting Δ (Supplementary Fig. S5c), and not κ. In this case, we used the measured refractive index of the silicon as an input in the simulation.

We experimentally demonstrated the critically coupled absorption from the flat band meta-optics. The presence of a flat band does not affect much while changing the Δ from 0.05 to 0.20, but we could clearly observe decreasing Q factors from the linewidth of the resonance (Supplementary Fig. S5d). We adjusted Δ from 0.05 to 0.35, and measured energy-momentum spectra in both reflection and transmission configuration. Then, extracted the Δ-dependent spectra at Λ-point (Supplementary Fig. S5, e and f), and sum of the reflection and transmission spectra is well-matched with the simulation which is ~68% (70%) absorption in measurement (simulation) when Δ = 0.20 (Fig. 2g and Supplementary Fig. S5c).

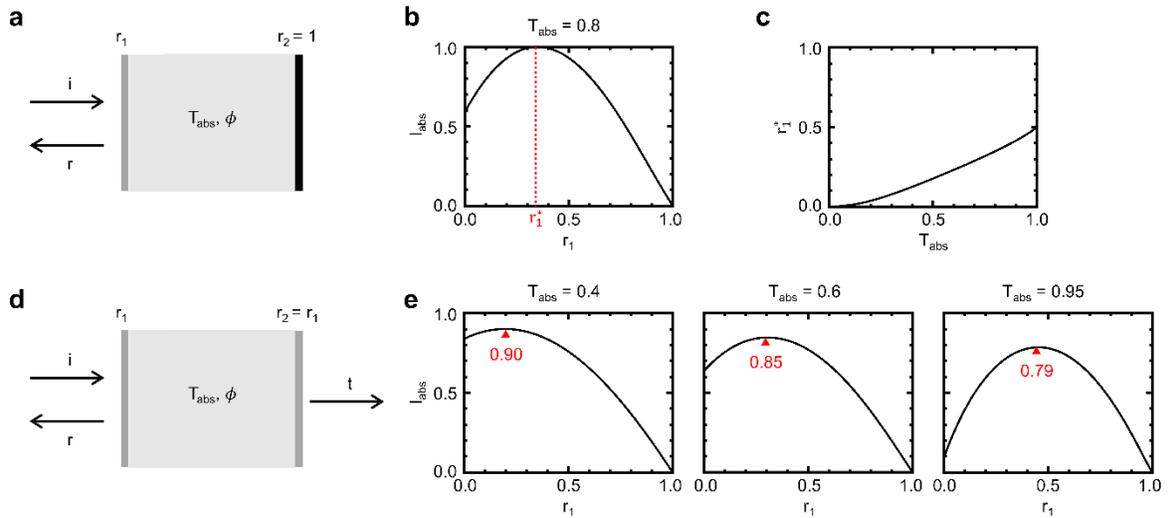

**Supplementary Figure S4. Critically-coupled absorption from a weak absorber and an optical resonator. a**, Schematic of a weak absorber placed in between two reflective lenses, where $r_1$, $r_2 = 1$, $T_{abs}$, and $\phi$ represent the reflectivity of the first and second mirrors, transmittance and phase delay along the weak absorber, respectively. **b**, Amount of light absorption with respect to $r_1$ when $T_{abs} = 0.8$ and $r_2 = 1$ ($r_1^*$ represents the optimized $r_1$ for maximizing $I_{abs}$, unity in this case). **c**, Optimized $r_1^*$ with respect to the $T_{abs}$ when $r_2 = 1$. **d**, Schematic of a weak absorber placed in between two reflective lenses, where $r_1 = r_2$. **e**, Amount of light absorption with respect to $r_1$ when $r_1 = r_2$ and $T_{abs} = 0.4$, 0.6, and 0.95, respectively. The maximum $I_{abs}$ is 0.90, 0.85, and 0.79 when $T_{abs} = 0.4$, 0.6, and 0.95, respectively, and there is a tendency that as $T_{abs}$ increases, $r_1^*$ also increases while $I_{abs}$ decreases.

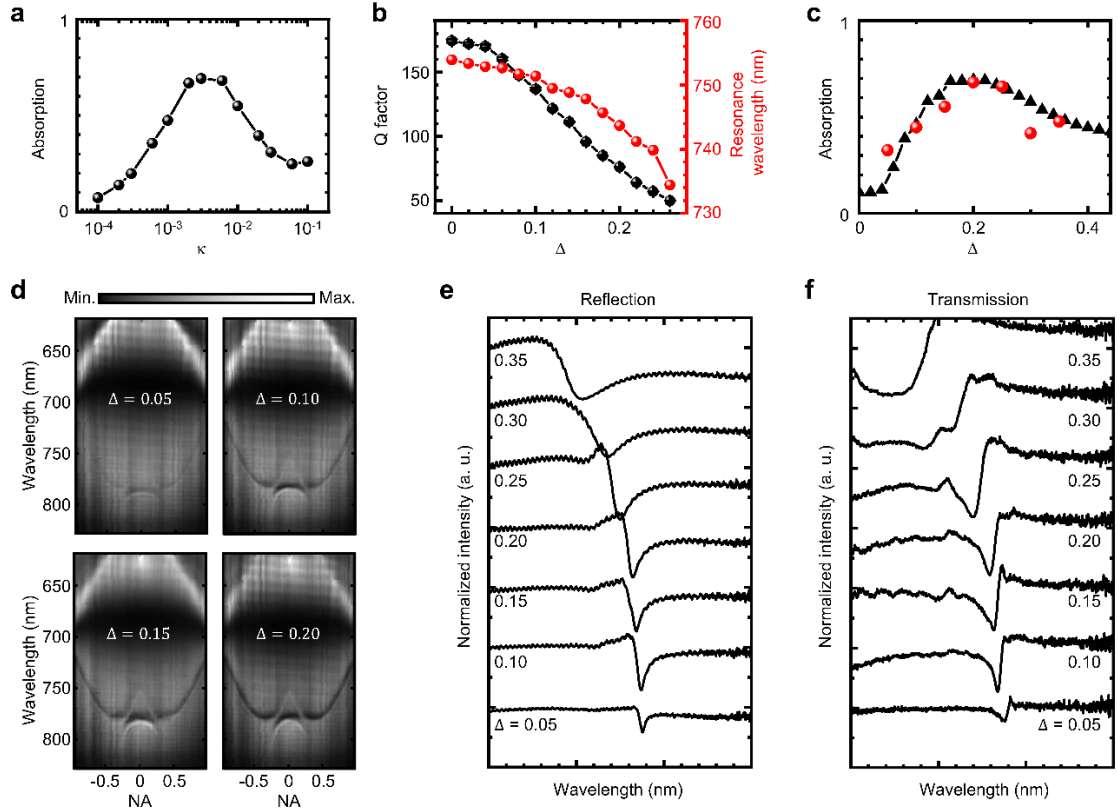

**Supplementary Figure S5. Critically-coupled absorption of the flat band meta-optics. a**, Simulated maximum absorption from the flat band meta-optics according to the κ of silicon using FDTD simulation. We used a model refractive index, $3.7 + i\kappa$, for silicon in FDTD simulation, where the maximum $I_{abs}$ of ~0.70 is achieved when κ is ~0.003. **b**, Simulated Q factor and resonance wavelength of the flat band meta-optics depending on Δ. As Δ increases, Q factor decreases, and the resonance wavelength is blue-shifted. **c**, Simulated (black) and measured (red) maximum absorption from the flat band meta-optics according to the Δ. **d**, Measured energy-momentum spectra of the flat band meta-optics with varying Δ, while the other parameters remain the same, in reflection configuration. **e**, **f**, Normalized reflection and transmission spectra of the flat band meta-optics at gamma-point according to the varying Δ from 0.05 to 0.35.

## III. Flat band meta-optics for efficient photodiode

We implemented the flat band meta-optic as a photodetector as it can absorb wide angular range of incident light with high efficiency at the same time, thanks to the wide angular acceptance of the flat band resonance. Hence, we can collect the light with a large aperture and focus on to the compact photodiode. In addition, the area where the nanofabrication is needed can be minimized as we can keep the meta-optics region small, which reduces the fabrication difficulties and enables mass production. Supplementary Fig. S6 represents the schematic procedure of how we fabricated the lateral p-i-n photodiode from the silicon flat band meta-optics.

We focused the light input into the photodiode and measured the power dependent photocurrent using the setup shown in Supplementary Fig. S7. Under the reverse bias voltage of −2V, we could clearly see the linear proportional relationship between the incident light power and photocurrent where the slope and y-intersects of the photocurrent at zero power represent a photon-to-current conversion efficiency and dark current, respectively. We plotted the raw data and linear plots of the power-dependent photocurrent for the flat band meta-optics photodiode for the wavelength from 765 to 799 nm with TE-polarization (Supplementary Fig. S8). We extracted the NA of the focused light from the focal spot size which corresponds to the Airy function. The FWHM of the beam size is $3.64 \pm 0.03$ μm, which corresponds to an NA of $0.132 \pm 0.001$. At the measured wavelength range from 765 to 799 nm, the extracted dark current remains $5.0 \pm 0.5$ nA, while the conversion efficiency is greatly enhanced near the resonance wavelength of the flat band (Supplementary Fig. S9).

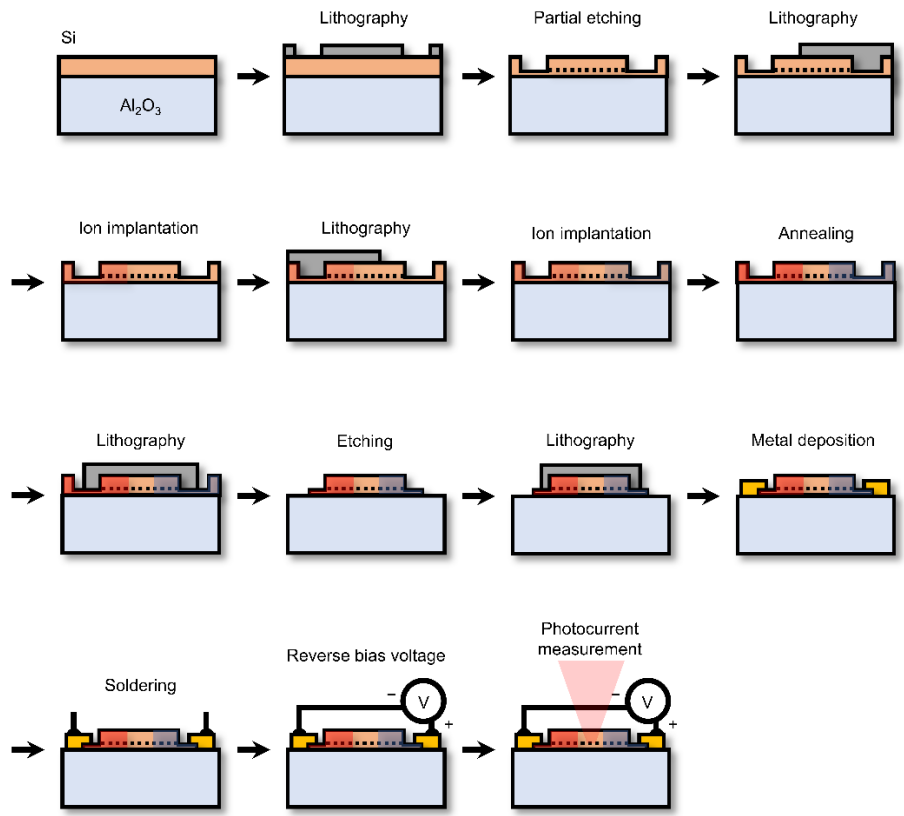

**Supplementary Figure S6. Schematic procedure for creating a lateral p-i-n photodiode from the silicon, flat band meta-optics.** We first created a photonic flat band from the crystalline-silicon on sapphire wafer, where the thickness of the silicon is about 230 nm. First, we spin-coated ZEP-520A resist on top of the wafer, then baked it at 180°C for 3 min. Then, we spin-coated electron discharger on it, did electron-beam-lithography, and developed with amyl acetate. We etched the silicon with mixture of $SF_6$ and $C_4F_8$ under $N_2$ atmosphere and remove the resist with 1-methyl-2-pyrrolidone. Then, we could confirm whether the meta-optics with photonic flat band was created by measuring the energy-momentum spectrum. If so, we did two sequences of lithography and ion implantation of with phosphorous (boron) ions for p- (n-) doping, respectively, and created create a lateral p-i-n photodiode. For the ion implantations, we used 2.00 $\times 10^{15}$ ions/cm$^2$ doses for both phosphorous and boron doping but with different acceleration voltage: 40 kV for phosphorous and 14 kV for boron. Then we anneal the ion implanted sample at 100°C for 10 min to activate the doping in the furnace. In order to operate the lateral p-i-n photodiode, we need to isolate the devices from the entire silicon layers. So, we patterned and etched all the unnecessary silicon layers to isolate all the p-i-n devices each other. In this case, we made 14 devices with different fill factors on a single chip in order to compensate the uncertain control of the partial etching depth, and made a metal contacts for all the devices at the same time. Two of those 14 devices are unetched and partially etched devices without any patterns for the references. Another optical lithography and electron-beam-evaporation followed by the liftoff process is done to create Ti/Au metal pads. Then we soldered two wires on each pad to stably apply bias voltage and measure the photocurrent. Lastly, we put the sample on the optical setup to selectively illuminate the focused light with an objective lens and measure the photocurrent under the reverse bias voltage.

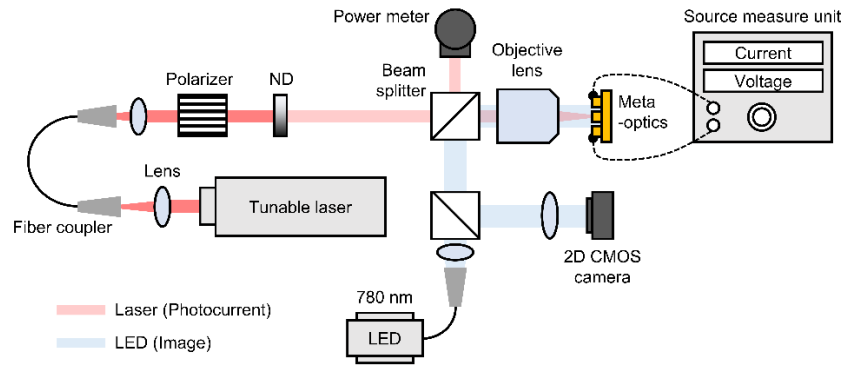

**Supplementary Figure S7. Schematic of the experimental setup for meta-optics photodiode characterization.** Tunable laser and polarizer are used to characterize both wavelength- and polarization-dependent response of the flat band meta-optics photodiode. We used an objective lens for both focusing the laser and also the imaging because we needed to observe whether the focused light is illuminating the appropriate device and the active region. A light-emitting diode (LED) of 780 nm wavelength and two-dimensional (2D) CMOS camera are used for imaging the device and alignment. While irradiating the device with focused light, we measured the photocurrent by connecting wires to both p- and n-contact metal pads and applying a reverse bias voltage of −2 V. With a neutral density (ND) filter, we can adjust the power of the irradiating light, and we simultaneously measure the incident light power and the photocurrent with a highly sensitive power meter and source measure unit, respectively.

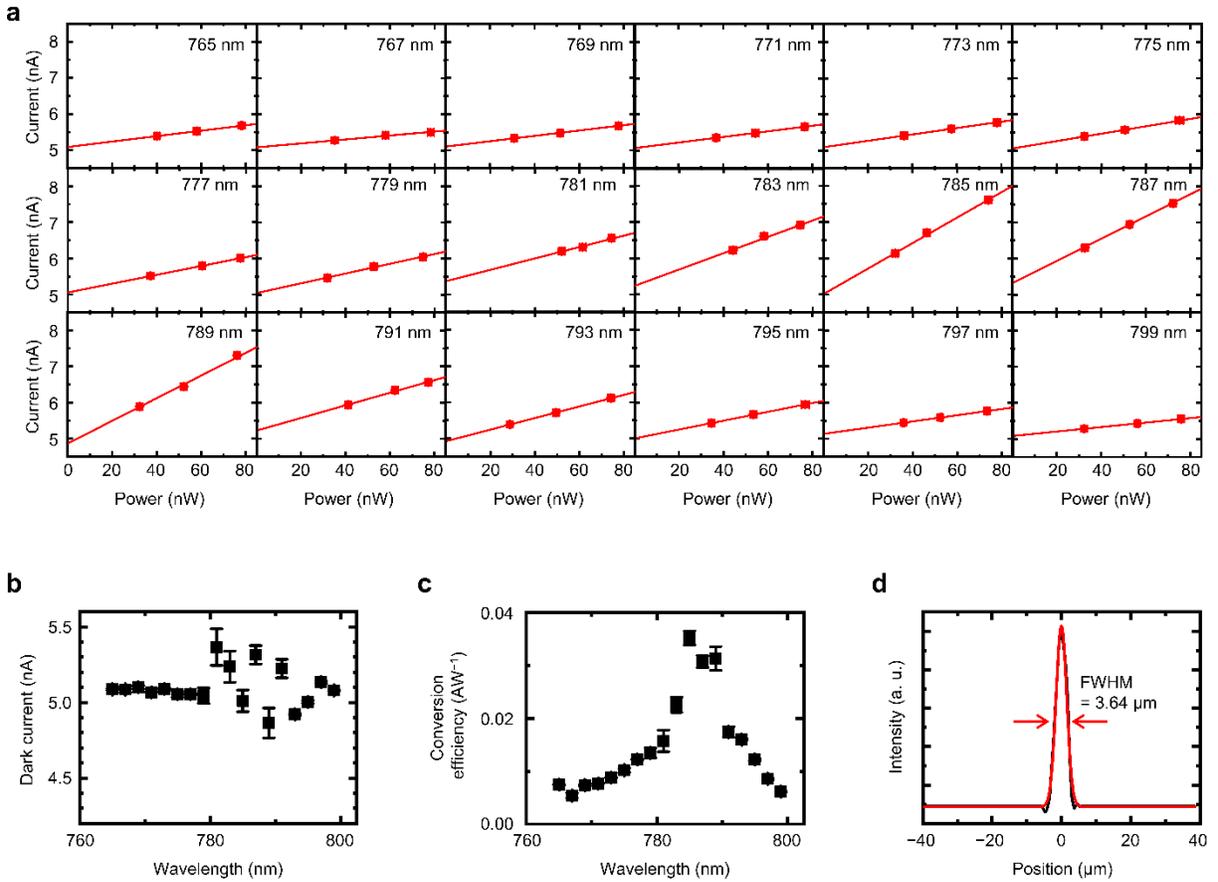

**Supplementary Figure S8. Wavelength dependent responsivity of the flat band photodiode for focused light with NA 0.13. a**, Power dependent photocurrent at the wavelength from 765 to 799 nm. The light is linearly polarized in TE-mode. **b**, **c**, Wavelength dependent dark current and conversion efficiency extracted from the linear functional fitting. Y-axis intercepts and the slopes of the linear functional fitting represents the dark current and photon-to-current conversion efficiency, respectively. The flat band meta-optics has its resonance wavelength ~785 nm. **d**, Line-profile of the focused light, and Gaussian functional fitting with full-width at half maximum (FWHM) value of 3.64 μm.

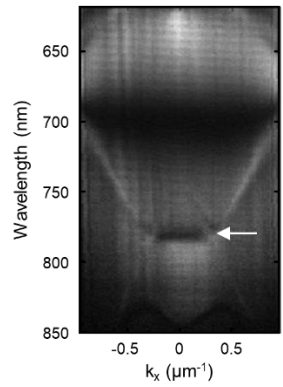

**Supplementary Figure S9. Energy-momentum spectrum of the flat-band photodiode in reflection configuration.** White arrow indicates the flat band, and the resonance is at the wavelength of about 785 nm.

## IV. Dimensional characteristics of the flat band

### 4.1. NA-dependent absorption of the flat band meta-optics

When we compare the photon-to-current conversion efficiency between the PDs with flat band meta-optics, non-flat band meta-optics, and unpatterned silicon film, we considered the light focused with an NA of 0.13. However, the enhancement factor of conversion efficiency strongly depends on the NA or working distance (WD) at fixed aperture size, of the system (Supplementary Fig. S10a). and we expect a higher enhancement from a flat band-based PD with a lower NA or long-WD. This peculiarity arises from the one-dimensionality of our flat band meta-optics, while using 2D rotationally symmetric lenses to focus light. Since we used a 1D flat band, cavity enhancement occurs only along one axis of the two-dimensional (2D) k-space of the incident light, as shown in Supplementary Fig. S10b; therefore, the ratio of cavity enhancement in k-space relative to all incident k-vectors is inversely proportional to the NA. The 10.3-fold enhancement was obtained by using a focusing lens with NA 0.13, and we expect a larger enhancement with a focusing lens of smaller NA (Supplementary Fig. S10c). In this regard, a trade-off arises between the performance and volume of the system due to the 1D characteristic of the flat band meta-optics.

To achieve both high performance and small volume on a system level, we ideally require a 2D flat band covering the entire k-space. In that case, all angles of incident light match the 2D flat band and undergo critically coupled absorption (Supplementary Fig. S10b). For this purpose, we adopt a rotationally symmetric, bullseye, structure to extend the flat band from 1D to 2D. This circular structure satisfies the configuration of a 1D flat band in all lateral directions passing through the center of the structure (Supplementary Fig. S11a). However, since photons perceive structures in a wavelength-scale volume, we needed to optimize the 2D bullseye structure transforming from a 1D flat band structure using two parameters: effective width of the cut line and the radius of the innermost circle (see details in the next Section 4.2). We further note that the 2D bullseye structure introduces a relationship between the k-vectors and polarization directions. As depicted in Supplementary Fig. S11a, each k-vector only fully matches the 1D flat band condition at the center of the Bullseye structure, but all different k-vectors need to match the orthogonal polarization directions. Therefore, not 100% of the incident light, but only 50% matches the flatband due to polarization selection.

We simulated the NA-dependent absorption of the 1D flat band and the 2D bullseye structures using FDTD (Supplementary Fig. S11b). We simulated the absorption of linearly-polarized light focused onto a 1D flat band meta-optic or the center of a 2D bullseye structure, as function of the focusing NA. From FDTD simulation, the 1D flat band structure shows higher absorption, up to 70%, for low-NA focused light, which is well-matched with our experimental result (Fig. 3, f and g in manuscript). However, as the NA of incoming light increases up to 0.35, the absorption decreases rapidly. On the other hand, the 2D bullseye structure exhibits ~35% absorption at low NA, which is half the 1D flat band due to polarization selection. But as the NA of incoming focused light increases, absorption decreases more slowly and eventually exceeds the 1D flat band at NA of 0.30 (Supplementary Fig. S11c).

As a result, we can expect higher absorption efficiency from the 1D flat band at linearly polarized and low-NA focused light inputs, while the 2D bullseye structures can be expected to have higher absorption at unpolarized or highly-focused light inputs. Additionally, the 2D bullseye structure is not free of alignment as the incoming light needs to be focused at the center, whereas the 1D flat band structure produces the same results regardless of the position where the light is irradiated. The measured energy-momentum spectra of the 2D bullseye structures are shown in Supplementary Fig. S12. We can see that the photonic band structures of 2D bullseye structures evolve from parabola to flatband and multivalley structures as the fill-factor increases, similar to the 1D flat band.

4.2. Design of 2D bullseye structure

We essentially brought the parameters – i.e. period, Δ, fill factor, and partial etch ratio – from the 1D flatband to 2D flatband. But as the light does have a volumetric response from the structure, we have to consider the effective width that light is experiencing along the gratings (Supplementary Fig. S13). There was no need to think about the effective width when the structure is 1D (Supplementary Fig. S13a); however, we have to consider the factor as the structure has 2D (Supplementary Fig. S13b).

We can think of the behavior of the light inside the medium as the function of effective refractive index of the medium. In this regard, the index and locations of each segment from 2D structure need to be identical to the 1D structure in order to expect the same effect as a photonic flatband. As shown in Supplementary Fig. S13b, if we convert the distance of each bar from 1D structure to the radius of the concentric bars of 2D structure, all the refractive segments are slightly moved toward the center due to the curved structure and the existence of the effective width, α. In this regard, we have to enlarge the radius ($r^*$) a little bit to compensate the center of the curved structure (Supplementary Fig. S13c). For each segment, the area of the 1D structure and 2D structure is as following ($r_2 > r_1 > 0$):

$$A_{1D}(r) = \alpha(r_2 - r_1),$$
$$A_{2D}(r) = \frac{1}{2}r_2^2\theta_2 - \frac{1}{2}r_1^2\theta_1 + \frac{1}{2}(r_2 - r_1)\alpha$$
$$= \frac{1}{2}r_2^2 \sin^{-1}\left(\frac{\alpha}{r_2}\right) - \frac{1}{2}r_2^2 \sin^{-1}\left(\frac{\alpha}{r_2}\right) + \frac{1}{2}(r_2 - r_1)\alpha$$
$$\cong \alpha(r_2 - r_1) - \frac{\alpha^3}{12}\left(\frac{1}{r_1} - \frac{1}{r_2}\right)$$
$$= \alpha(r_2 - r_1)\left(1 - \frac{\alpha^2}{12r_1r_2}\right)$$

It is clear that the $A_{1D} \neq A_{2D}$, so we need to modulate the radius for 2D bullseye structure to make them identical.

$$A_{1D}(r) = A_{2D}(r^*)$$
$$\alpha(r_2 - r_1) = \alpha(r_2^* - r_1^*)\left(1 - \frac{\alpha^2}{12r_1^*r_2^*}\right)$$

As we define the modulation parameter of X to be $r^* = Xr$, then the equation becomes as following:

$$\frac{1}{X} \equiv \frac{r_2 - r_1}{r_2^* - r_1^*} = 1 - \frac{\alpha^2}{12r_1^*r_2^*} = 1 - \frac{\alpha^2}{12r_1r_2}\frac{1}{X^2} \cong 1 - \frac{\alpha^2}{12r^2}\frac{1}{X^2}$$

By multiplying $X^2$ in both sides of the equation, we got:

$$X^2 - X - \frac{\alpha^2}{12r^2} = 0$$
$$X = \frac{1}{2} + \sqrt{\frac{1}{4} + \frac{\alpha^2}{12r^2}}$$

As a result, we can see the relationship between r and $r^*$ as following:

$$r^* = r\left(\frac{1}{2} + \sqrt{\frac{1}{4} + \frac{\alpha^2}{12r^2}}\right) > r$$

For verification, we can see that $r^*$ is greater than r, and $r^* = r$ as the effective width, α becomes zero or r becomes infinite which all makes sense.

In this regard, we ran FDTD simulations with two modulation parameters – (1) Radius of the inner most center circle, and (2) An effective width. And we can design four different configurations for the bullseye structure depend on the central meta atoms as shown in Supplementary Fig. S13d. And we can see that the upper two configurations which has a larger area of absorptive medium at the center has exactly the same absorption power larger than the case of lowers which has smaller inner circles.

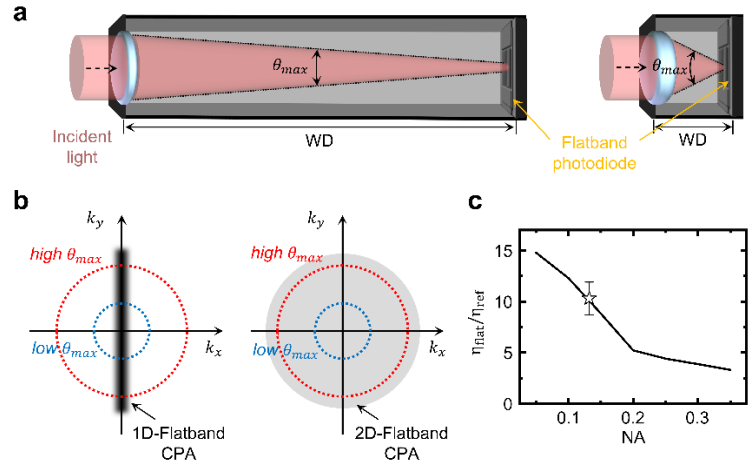

**Supplementary Figure S10. NA-dependent absorption in the 1D flat band**. **a**, Schematics of the flat band-based PD with different WD of the focusing lenses. **b**, Regions of incident light with high- (red) and low-NA (blue) focusing lens and areas of 1D and 2D flat band in lateral k-space. **c**, Simulated absorption power enhancement of 1D flat band compared to the 230-nm-thick silicon film depend on NA of the focused input light. A star symbol indicates the measured value at NA of 0.13.

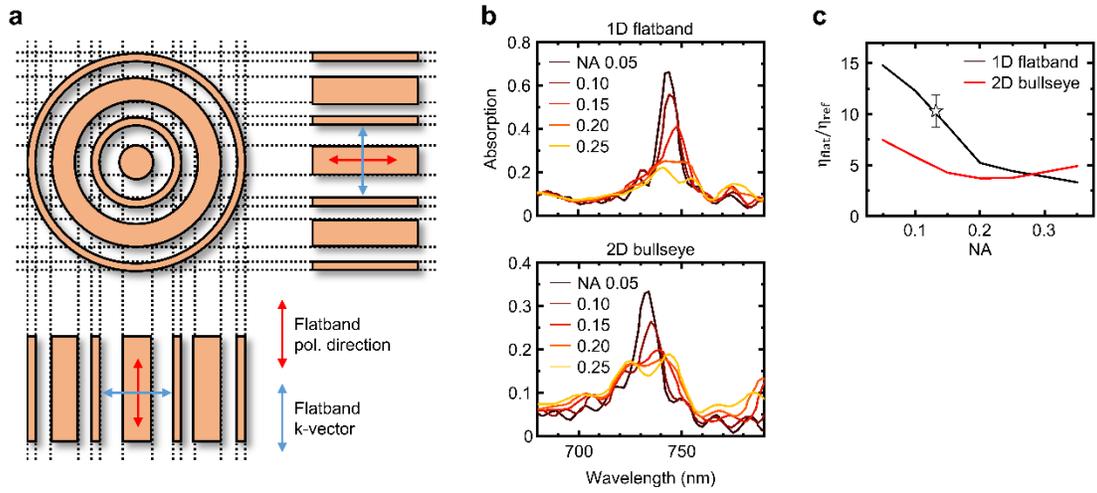

**Supplementary Figure S11. NA dependent absorption in 2D bullseye structure compared to 1D flat band. a**, Schematic of the 2D Bullseye structure and corresponding k-vectors and polarization directions at the resonance. **b**, Simulated absorption spectra of 1D flat band and 2D bullseye structures with respect to the NA of the focused incident light. **c**, Simulated absorption power enhancement of 1D flat band (black) and 2D bullseye (red) structures compared to the 230-nm-thick silicon film depend on NA of the focused input light. A star symbol indicates the measured value at NA of 0.13.

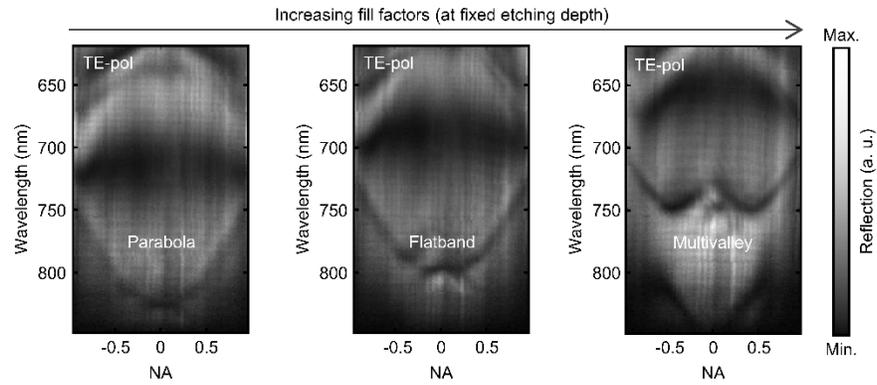

**Supplementary Figure S12. Energy-momentum spectra of the 2D bullseye structure.** Measured energy-momentum spectra of the 2D bullseye structure with varying fill factors. The photonic band structures are tuned from parabola to flat band, and multivalley.

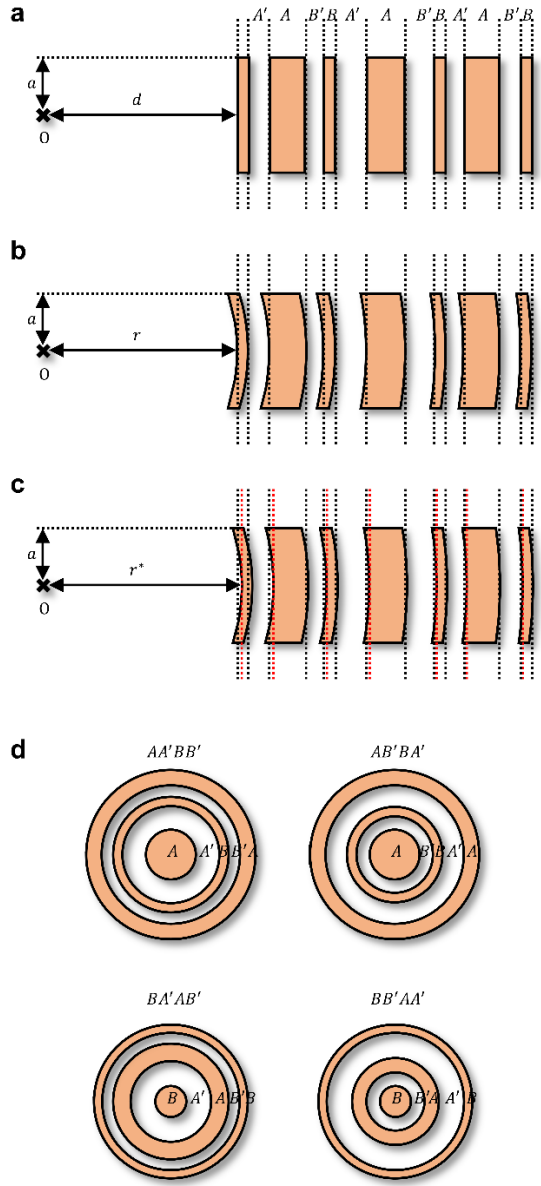

**Supplementary Figure S13. Parameters to convert 1D flatband to 2D bullseye structure. a**, Schematics of the 1D flatband. **b**, **c**, Schematics of the 2D flatband before and after considering the curved shapes. **d**, Four different configurations of the 2D bullseye structure.